\documentclass[12pt,a4paper,final]{iopart}

\usepackage{amstext}
\usepackage{subfigure}
\usepackage{iopams}  
\usepackage{graphicx}
\usepackage{epstopdf}
\usepackage[breaklinks=true,colorlinks=true,linkcolor=blue,urlcolor=blue,citecolor=blue]{hyperref}

\newcommand{\MM}[1]{{\color{green} #1}}

\usepackage{lineno}
\usepackage{threeparttable}
\usepackage{color}
\usepackage[section]{placeins}
\usepackage{comment}
\newcommand{\comm}[1]{}
\usepackage{epstopdf}
\usepackage{float}
\usepackage{multirow}
\usepackage{subfigure}
\usepackage{hyperref}
\usepackage{enumerate}
\modulolinenumbers[5]

\begin{document}
	

\article[ A comment on "The Computational 2D Materials Database..."]{COMMENT}{A comment on "The Computational 2D Materials Database: high-throughput modeling and discovery of atomically thin crystals"}

\author{Marcin Ma\'zdziarz}
\address{Institute of Fundamental Technological Research Polish Academy of Sciences,
Pawi\'nskiego 5B, 02-106 Warsaw, Poland}
\ead{mmazdz@ippt.pan.pl}

\begin{abstract}
{Recently, Sten Haastrup, Mikkel Strange, Mohnish Pandey, Thorsten Deilmann, Per S Schmidt, Nicki F Hinsche, Morten N
Gjerding, Daniele Torelli, Peter M Larsen, Anders C Riis-Jensen, Jakob Gath, Karsten W Jacobsen, Jens
J$\o$rgen Mortensen, Thomas Olsen and Kristian S Thygesen [2D Mater. 5 (2018) 042002] \cite{Haastrup2018} introduced the Computational 2D Materials Database (C2DB), which organises a variety of structural, thermodynamic, elastic, electronic, magnetic, and optical properties of around 1500 two-dimensional materials distributed over more than 30 different crystal structures. 
Unfortunately, the work contains serious and fundamental flaws in the field of elasticity and mechanical stability tests what makes it unreliable.}
\end{abstract}

\vspace{2pc}


\section*{}
\label{sec:Int}

In \cite[figure 1]{Haastrup2018} the workflow used to calculate the structure and properties of the materials in C2DB the authors stated that the dynamical stability condition for a structure is not satisfied when elastic constants are negative. Unfortunately, it is an incorrect condition.
Moreover, in \cite[equation (3)]{Haastrup2018} the authors, for reasons difficult to understand, disregarded shear deformations and calculated only the planar elastic stiffness coefficients $C_{11}$, $C_{22}$, and $C_{12}$, what makes the aforementioned analysis incomplete and insufficient. \comm{\MM}
{In addition, even these calculated coefficients in C2DB are erroneous, i.e. the stiffness tensor does not have a proper symmetry resulting from the symmetry of the crystal.} 

In order to explain what the problem is, some facts from the theory of 2D linear elasticity and elastic stability analysis should be recalled.

The generalised Hooke's law is the linear strain-stress tensor relation:
\begin{equation}\label{hl}
\sigma_{ij}=C_{ijkl}\varepsilon_{kl}\rightarrow \boldsymbol{\sigma}=\mathbb{C}\boldsymbol{\varepsilon},
\end{equation}
where $\boldsymbol{\sigma}$ is the second-rank Cauchy stress tensor,
$\mathbb{C}$ is the fourth-rank anisotropic elastic stiffness tensor and $\boldsymbol{\varepsilon}$ is the second-rank small strain tensor ($i,j,k$=1,2,3 for 3D and $i,j,k$=1,2 for 2D problems), from Einstein
summation convention repeated indices are implicitly summed over.

From the symmetry of $\boldsymbol{\sigma}$ and $\boldsymbol{\varepsilon}$ it follows that
\begin{equation}\label{cs1}
C_{ijkl}=C_{jikl}=C_{ijlk},
\end{equation}

and from the thermodynamic requirement of existing of a strain energy density function $U(\boldsymbol{\varepsilon})$ (hyperelastic material) \cite{Carroll09} such that
\begin{equation}\label{se}
U=\frac{1}{2}\frac{\partial^2 U}{\partial \varepsilon_{ij} \partial \varepsilon_{kl}}\bigg|_{\boldsymbol{\varepsilon}=0}\varepsilon_{ij}\varepsilon_{kl} = \frac{1}{2}C_{ijkl}\varepsilon_{ij}\varepsilon_{kl}\rightarrow U=\frac{1}{2}\boldsymbol{\varepsilon}\mathbb{C}\boldsymbol{\varepsilon},
\end{equation}

additionally
\begin{equation}\label{cs2}
C_{ijkl}=C_{klij},
\end{equation}
and hence number of independent components of four-rank $C_{ijkl}$ reduces to 21 in 3D \cite{hetnarski2010mathematical} and to 6 in 2D \cite{Blinowski1996, He1996}.
In relations (\ref{hl}) and (\ref{se}) the \textit{fourth-rank tensor} notation, employing fourth-rank Cartesian tensor in three or two dimensions, is used. Also different notations for the generalized Hooke's law, relation (\ref{hl}), are in use. The most popular is a non-tensorial \textit{Voigt} notation that employs in 2D 3x3 matrix: 
\begin{eqnarray}
\centering
\left[
\begin{array}{c}
{\sigma_{11}}\\
{\sigma_{22}}\\
{\sigma_{12}} \\
\end{array}
\right]=\left[
\begin{array}{ccc}
{C_{1111}} & {C_{1122}} & {C_{1112}} \\
{C_{1122}} & {C_{2222}} & {C_{2212}} \\
{C_{1112}} & {C_{2212}} & {C_{1212}} \\
\end{array}
\right] \left[
\begin{array}{c}
{\varepsilon_{11}} \\
{\varepsilon_{22}} \\
2{\varepsilon_{12}} \\
\end{array}
\right],\ 
\label{eqn:HookVoight}
\end{eqnarray}

or
\begin{eqnarray}
\centering
\left[
\begin{array}{c}
{\hat{\sigma}_{1}}\\
{\hat{\sigma}_{2}}\\
{\hat{\sigma}_{3}} \\
\end{array}
\right]=\left[
\begin{array}{ccc}
{\hat{C}_{11}} & {\hat{C}_{12}} & {\hat{C}_{13}} \\
{\hat{C}_{12}} & {\hat{C}_{22}} & {\hat{C}_{23}} \\
{\hat{C}_{13}} & {\hat{C}_{23}} & {\hat{C}_{33}} \\
\end{array}
\right] \left[
\begin{array}{c}
{\hat{\varepsilon}_{1}} \\
{\hat{\varepsilon}_{2}} \\
{\hat{\gamma}_{3}} \\
\end{array}
\right]\rightarrow \boldsymbol{\hat{\sigma}}=\boldsymbol{\hat{c}}\boldsymbol{\hat{\varepsilon}}.\ 
\label{eqn:HookVoight2}
\end{eqnarray}
The less popular is a \textit{second-rank tensor}, called also \textit{orthonormal} or \textit{Mandel}, notation: 
\begin{eqnarray}
\centering
\left[
\begin{array}{c}
{\sigma_{11}}\\
{\sigma_{22}}\\
\sqrt{2} {\sigma_{12}} \\
\end{array}
\right]=\left[
\begin{array}{ccc}
{C_{1111}} & {C_{1122}} & \sqrt{2}{C_{1112}} \\
{C_{1122}} & {C_{2222}} & \sqrt{2}{C_{2212}} \\
\sqrt{2}{C_{1112}} & \sqrt{2}{C_{2212}} & 2{C_{1212}} \\
\end{array}
\right] \left[
\begin{array}{c}
{\varepsilon_{11}} \\
{\varepsilon_{22}} \\
\sqrt{2} {\varepsilon_{12}} \\
\end{array}
\right],\ 
\label{eqn:HookOrthonormal}
\end{eqnarray}

or
\begin{eqnarray}
\centering
\left[
\begin{array}{c}
{{\sigma}_{1}}\\
{{\sigma}_{2}}\\
{{\sigma}_{3}} \\
\end{array}
\right]=\left[
\begin{array}{ccc}
{{C}_{11}} & {{C}_{12}} & {{C}_{13}} \\
{{C}_{12}} & {{C}_{22}} & {{C}_{23}} \\
{{C}_{13}} & {{C}_{23}} & {{C}_{33}} \\
\end{array}
\right] \left[
\begin{array}{c}
{{\varepsilon}_{1}} \\
{{\varepsilon}_{2}} \\
{{\varepsilon}_{3}} \\
\end{array}
\right]\rightarrow \boldsymbol{\sigma}=\boldsymbol{{c}}\boldsymbol{\varepsilon}.\ 
\label{eqn:HookOrthonormal2}
\end{eqnarray}

The difference between \textit{Voigt} and \textit{second-rank tensor} notation is not only by factors of 2 and its square root but is more fundamental. In the \textit{Voigt} notation, the elements of matrix $\boldsymbol{\hat{c}}$ in equation (\ref{eqn:HookVoight2}) are not the elements of a second-rank tensor, whereas in the \textit{second-rank tensor} notation the elements of $\boldsymbol{{c}}$ in equation (\ref{eqn:HookOrthonormal2}) are the elements of a {second-rank tensor} in six dimensions for 3D and three dimensions for 2D problems. The \textit{fourth-rank tensor} notation (\ref{hl}) and \textit{second-rank tensor} notation (\ref{eqn:HookOrthonormal2}) are tensorially equivalent \cite{MEHRABADI1990, Blinowski1996}. 

In two-dimensional space, there are five different cell lattice types: 
\begin{enumerate}[I.]
	\item Oblique (parallelogram) (a$\neq$b, $\measuredangle\neq$90$^{\circ}$) \label{itm:1}, \label{itm:o}
	\item Rectangular (a$\neq$b, $\measuredangle$=90$^{\circ}$) \label{itm:2},\label{itm:r}
	\item Centered rectangular or diamond (a$\neq$b, $\measuredangle$=90$^{\circ}$) \label{itm:3},\label{itm:cr}
	\item Square (a=b, $\measuredangle$=90$^{\circ}$) \label{itm:4},\label{itm:s}
	\item Rhombic or hexagonal (a=b, $\measuredangle$=120$^{\circ}$) \label{itm:5}.\label{itm:h}
	\label{lt}
\end{enumerate}

It is clear that symmetry aspects are important in the study of physical phenomena.
\comm{It is known that symmetry considerations play an important role in the study of physical phenomena.} From \textit{symmetry principle: if a crystal is invariant with respect to certain symmetry elements, any of its physical properties must also be invariant with respect to the same symmetry elements} and \textit{Curie laws}, it results that the symmetries of the physical properties of the material may not be lower than the symmetry of the crystal, but may be higher \cite{nye1985physical, Dimitrienko2002}.
\comm{it arises that the symmetries in the physical properties of a material cannot be smaller than the symmetry of the crystal, but can be bigger.}

The symmetry classification of linear elastic materials is not related to crystallography. This is due to the properties of fourth-rank Euclidean symmetric tensors (from the linearity of  phenomenological Hooke's law and the properties of two, three-dimensional Euclidean space)\cite{Mazdziarz15}. For 3D linear hyperelastic materials, there are eight classes of symmetry and four classes of symmetry for 2D \cite{Mazdziarz15, Blinowski1996}. 

\comm{It must also be remembered that classification of linear-elastic materials due to the symmetries leads only to eight classes of symmetry, while in crystallography are distinguished 32 crystallographic point groups and 7 \textit{Curie groups}. 
	Distribution of linear-elastic materials by symmetry has nothing to do with crystallography. It results from the properties of fourth order Euclidean symmetric tensors (from linearity of Hooke's relation and properties of two, three-dimensional Euclidean space).}


Necessary and sufficient elastic stability conditions, also called Born stability
conditions, in various 3D crystal systems are gathered in \cite{Mouhat2014}, but from my best knowledge, there is no such work for 2D crystal systems. 

In general, the unstressed crystalline structure is stable with no external loads and in the harmonic approximation, if and only if two independent conditions are fulfilled:
\begin{enumerate}[1.]
	\item All its phonon modes have positive frequencies $\boldsymbol{\omega}$ for all wave vectors $\boldsymbol{q}$ (dynamical stability):
	\begin{equation}\label{ds}
	\boldsymbol{\omega^2(q)}>0,
	\end{equation}
	\item The strain energy density function, given by the quadratic form (\ref{se}), is always
	positive (elastic stability):
	\begin{equation}\label{es}
	U(\boldsymbol{\varepsilon})>0, \forall{\boldsymbol{\varepsilon}}\neq 0.
	\end{equation}
\end{enumerate}
It is worth pointing out that some authors incorrectly identify elastic stability (\ref{es}) with dynamic stability (\ref{ds}) for the long wave limit (i.e. for vanishing wavevectors $\boldsymbol{q}\rightarrow$0) \cite{Grimvall2012, Rehak2012}. In the mathematical elasticity this phonon condition is called strong ellipticity and does not imply positive definiteness of the strain energy density function (\ref{se}), but the opposite implication occurs \cite{hetnarski2010mathematical}.

It would be quite difficult to check the positive definiteness of the quadratic form (\ref{es}) directly and it can, therefore, be replaced by equivalent easier conditions \cite{Mouhat2014}:
\begin{enumerate}[1.]
\item All eigenvalues of tensor $\boldsymbol{{c}}$ in \textit{second-rank tensor} notation (\ref{eqn:HookOrthonormal2}) are positive,\

or 
\item All the leading principal minors of tensor $\boldsymbol{{c}}$ in (\ref{eqn:HookOrthonormal2}) (determinants of
its upper-left k by k submatrix) are positive (Sylvester’s criterion).
\end{enumerate}




After this theoretical introduction we can give the form of elastic stiffness tensor $\boldsymbol{{c}}$ in the \textit{second-rank tensor} notation (\ref{eqn:HookOrthonormal2}) and the necessary and sufficient elastic stability conditions (\ref{es}) for all four classes of symmetry for 2D hyperelastic materials.

\begin{enumerate}[1.]
\item Full symmetry (isotropy) $\rightarrow$ Hexagonal lattice {(\ref{itm:h})} \newline  
(2 elastic constants)
\label{itm:iso}
\begin{eqnarray}
\centering
{C}_{IJ}
\rightarrow\left[
\begin{array}{ccc}
{{C}_{11}} & {{C}_{12}} & 0 \\
{{C}_{12}} & {{C}_{11}} & 0 \\
0 & 0 & C_{11}-C_{12} \\
\end{array}
\right], 
\label{eqn:Hisotropy}
\end{eqnarray}
$C_{11}>0$ \& $C_{11} > \left|C_{12}\right|$ or
$\lambda_I=(C_{11} + C_{12})>0$ \& $\lambda_{II}=(C_{11} - C_{12})>0$.
\item Symmetry of a square, (tetragonal)$\rightarrow$ Square lattice {(\ref{itm:s})}\newline
(3 elastic constants)
\label{itm:sq}
\begin{eqnarray}
\centering
{C}_{IJ}
\rightarrow\left[
\begin{array}{ccc}
{{C}_{11}} & {{C}_{12}} & 0 \\
{{C}_{12}} & {{C}_{11}} & 0 \\
0 & 0 & C_{33} \\
\end{array}
\right], 
\label{eqn:Htetragonal}
\end{eqnarray}
$C_{11}>0$ \& $C_{33}>0$ \& $C_{11} > \left|C_{12}\right|$  or
$\lambda_I=(C_{11} + C_{12})>0$ \& $\lambda_{II}=(C_{11} - C_{12})>0$ \& $\lambda_{III}=C_{33}>0$.
\item Symmetry of a rectangle, (orthotropy)$\rightarrow$ Rectangular {(\ref{itm:r})} \& Centered rectangular lattice {(\ref{itm:cr})} \newline
(4 elastic constants)
\label{itm:rec}
\begin{eqnarray}
\centering
{C}_{IJ}
\rightarrow\left[
\begin{array}{ccc}
{{C}_{11}} & {{C}_{12}} & 0 \\
{{C}_{12}} & {{C}_{22}} & 0 \\
0 & 0 & C_{33} \\
\end{array}
\right], 
\label{eqn:Horthotropy}
\end{eqnarray}
$C_{11}>0$ \& $C_{33}>0$ \& $C_{11}C_{22} > {C^2_{12}} $ or
$\lambda_I=\frac{1}{2} \left(C_{11} + C_{22}+\sqrt{4C^2_{12}-(C_{11}-C_{22})^2}\right)>0$ \& $\lambda_{II}=\frac{1}{2} \left(C_{11} + C_{22}-\sqrt{4C^2_{12}-(C_{11}-C_{22})^2}\right)>0$ \& $\lambda_{III}=C_{33}>0$. 
\item No symmetry (anisotropy) $\rightarrow$ Oblique lattice {(\ref{itm:o})}\newline
(6 elastic constants)
\label{itm:aniso}
\begin{eqnarray}
\centering
{C}_{IJ}
\rightarrow\left[
\begin{array}{ccc}
{{C}_{11}} & {{C}_{12}} & {{C}_{13}} \\
{{C}_{12}} & {{C}_{22}} & {{C}_{23}} \\
{{C}_{13}} & {{C}_{23}} & {{C}_{33}} \\
\end{array}
\right], 
\label{eqn:Hanisotropy}
\end{eqnarray}
$C_{11}>0$ \& $C_{11}C_{22} > {C^2_{12}}$ \& det(${C}_{IJ}$) > 0 or $\lambda_I>0$ \& $\lambda_{II}>0$ \& $\lambda_{III}>0$ (e.g. from the Cardano formula \cite{Itskov07}).
\end{enumerate}

The problem can arise if we find $C_{13}$ and/or  $C_{23}$ other than zero: it is hard
to say, in this case, if there is no symmetry at all or, possibly, we have chosen a
wrong axis \cite{Blinowski1996}. To avoid this it is recommended to check for all crystals the most general stability condition for anisotropy (\ref{eqn:Hanisotropy}). 

\comm{\MM}
{The above considerations are not only of a general nature, selected examples of erroneous stiffness tensors and incorrectly verified elastic stability can be found in the Computational 2D Materials Database (C2DB). \label{sec:C2DB} \newline
As it was written earlier, crystal symmetry implies symmetries of its physical properties, and hence the symmetries of tensors, e.g. the stiffness tensor. The conditions for elastic stability were given in equations (\ref{eqn:Hisotropy}--\ref{eqn:Hanisotropy}).\newline
For example, we can find in the C2DB database:
\begin{itemize}
	    \item \textbf{Au$_2$O$_2$}: https://cmrdb.fysik.dtu.dk/c2db/row/Au2O2-GaS-NM\newline
	Space group:P-6m2, $C_{11}$=86.93\,N/m, $C_{22}$=87.90\,N/m and $C_{12}$=103.62\,N/m\newline
	Because it is a Hexagonal lattice {(\ref{itm:h})} the stiffness tensor $\boldsymbol{{c}}$ must be isotropic (\ref{itm:iso}) and  $C_{11}$ must be equal to $C_{22}$. Although all calculated elastic constants are positive, the crystal is not elastically stable because not all required stability conditions (Eq.\ref{eqn:Hisotropy}) are fulfilled. 
		\item \textbf{Ta$_2$Se$_2$}: https://cmrdb.fysik.dtu.dk/c2db/row/Ta2Se2-GaS-FM\newline
	Space group:P-6m2, $C_{11}$=75.15\,N/m, $C_{22}$=75.81\,N/m and $C_{12}$=-45.67\,N/m\newline
	Because it is a Hexagonal lattice {(\ref{itm:h})} the stiffness tensor $\boldsymbol{{c}}$ must be isotropic (\ref{itm:iso}) and  $C_{11}$ must be equal to $C_{22}$. Although calculated elastic constant $C_{12}$ is negative, the crystal is elastically stable because all mandatory stability conditions (Eq.\ref{eqn:Hisotropy}) are satisfied. 
		\item \textbf{Re$_2$O$_2$}: https://cmrdb.fysik.dtu.dk/c2db/row/Re2O2-FeSe-NM\newline
	Space group:P4/nmm, $C_{11}$=17.70\,N/m, $C_{22}$=16.18\,N/m and $C_{12}$=239.42\,N/m\newline
	Because it is a Square lattice {(\ref{itm:s})} the stiffness tensor $\boldsymbol{{c}}$ must have symmetry of a square (\ref{itm:sq}) and  $C_{11}$ must be equal to $C_{22}$ (the difference here is more than 9\%). Although all calculated elastic constants are positive, the crystal is not elastically stable because not all stability requirements (Eq.\ref{eqn:Htetragonal}) are met.  
\end{itemize} 
}
\section*{Acknowledgments}
This work was partially supported by the National Science Centre (NCN -- Poland) Research Project: UMO-2016/21/B/ST8/02450.











\bibliographystyle{unsrt} 
\section*{References}
\bibliography{2DMaterials}
\end{document}